\documentclass[preprint,preprintnumbers,amsmath,amssymb]{revtex4}

\usepackage{graphicx}
\usepackage{dcolumn}
\usepackage{bm}

\begin{document}

\title{Information content of financial markets: a practical approach based on Bohmian quantum mechanics}
\author{F. Tahmasebi $^1$, S. Meskini $^1$, A. Namaki $^2$, G.R. Jafari $^1$ \thanks{Email: g\_jafari@sbu.ac.ir} $^\dag$\\
{\small $^1$ Department of Physics, Shahid Beheshti University,
G.C., Evin, Tehran 19839, Iran}\\
{\small $^2$ Department of Financial Management, Faculty of
Management, University of Tehran, Tehran, Iran} }

\begin{abstract}

The Bohmian quantum approach is implemented to analyze the financial markets. In this approach, there is a wave function that leads to a quantum potential. This potential can explain the relevance and entanglements of the agent's behaviors with the past. The light is shed by considering the relevance of the market conditions with the previous market conditions enabling the conversion of the local concepts to the global ones.
We have shown that there are two potential limits for each market. In essence, these potential limits act as a boundary which limits the return values inside it. By estimating the difference between these two limits in each market, it is found that the quantum potentials of the return time series in different time scales, possess a scaling behavior. The slopes of the scaling behaviors in mature, emerging and commodity markets show different patterns. The emerge market having a slope greater than 0.5, has a higher value compared to the corresponding values for the mature and commodity markets which is less than 0.5. The cut-off observed in the curve of the commodity market indicates the threshold for the efficiency of the global effects. While before the cut-off, local effects in the market are dominant, as in the case of the mature markets.
The findings could prove adequate for investors in different markets to invest in different time horizons.

\end{abstract}
\maketitle

\section{Introduction}

In the mainstream economy, the cornerstone for mathematical modeling
of financial markets is the efficient market hypothesis
(EMH)\cite{Campbell,Harrison}. A market is said to be efficient if
all the existing information has affected the prices and any new
random information influences the asset prices randomly
\cite{Campbell}. The EMH hypothesis is based on some assumptions
such as rationality of the investors. Empirical studies have shown
that the real prices do not completely follow a random walk which is
the criterion of EMH \cite{Mantegna}. Some methods such as
behavioral models, try to explain the real behavior of asset prices.
On the other hand, one of the behavioral aspects of markets is the
entanglement of their conditions that lead to converting the local
concepts to the global ones. For example, the today market
conditions have a dependency to the previous conditions. So,
researchers try to use new methods in order to model the behavior of
the markets that consider the coupling structure of them. One of the
latest and most powerful approaches for this purpose is
implementation of the concepts of quantum mechanics. The previous
studies have shown the usefulness of this approach for analyzing the
financial markets. Application of quantum formalism in Finance and
economy has been shown by  Seagal et.al \cite{Seagal} where they used
quantum mechanics  for option pricing. The work was completed by the
paper of Accardi et al \cite{Accardi}. Baaquie has used the quantum
concepts in different domains of
finance \cite{Baaquie,Baaquie1,Baaquie2,Baaquie3,Baaquie4,Baaquie5,Pedram}.

Haven has embedded the Black- Scholes option pricing model in a
quantum physics setting\cite{Haven,Haven1,Haven2,Haven3}.
 One of
the latest quantum formalisms for analyzing the financial markets is
using the concept of Bohmian
mechanics\cite{Choustova,Choustova1,Choustova2}.

In the Bohmian framework, there is a potential quantum term that
shows entanglements and interactions between particles very well.
So, this potential term is beneficiary for presenting the structure
of the coupled systems such as financial markets that have different
participant agents. Choustova \cite{Choustova,Choustova1,Choustova2}
presented a model for describing the structure of the financial
markets based on the Bohmian mechanics.

The model can describe the complexity of the financial markets by
considering the psychology of agents of these markets in an
information field. In essence, this model can explain the behavioral
aspects of markets by using the Bohmian quantum
mechanics \cite{Choustova,Choustova1,Choustova2,Choustova3}. In this paper, we
have applied the Chostova's proposed method in order to extract the
information content of financial return time series. We have used
different databases; covering securities from the Tehran stock
exchange (TSE) as an emerge market \cite{TSE}, Dow Jones Industrial Average
(DJIA30), Standard and Poor's 500 (SPX) indices from a mature
market, and gold \& west Texas intermediate oil prices as commodity
markets. We have analyzed the daily change of these markets from the
1st of Jan 1996 until 1st of Jan 2011.

The paper is organized as follows: In section II we explain the relation between the Bohmian mechanics and finance theory. In section III we present our findings before stating our conclusions in section IV.

\section{Finance and Bohmian Quantum Mechanics}

Bohemian quantum mechanics is a special point of view to quantum
theory with a causal interpretation. This view was first presented
by Louis de Broglie in 1927, which was later rediscovered by David
Bohm in 1952 \cite{Bohm,Bohm1}. In Bohmian mechanics, a system of
particles is described by a wave function that evolves according to
Schr$\ddot{o}$dinger's equation and a deterministic motion
part \cite{Bohm}. The essential part of Bohmian quantum is based on
two special assumptions names as the non-locality and the
deterministic trajectory. This deterministic trajectory means that
there is a guiding equation stemmed from Newton equation that
presents the actual positions of the particles. The non-locality
means that the behavior of any particle through the wave function
which depends on the whole configuration of the universe relates to
the other particles statues. This property is equivalent to the
notion of the entanglement which plays an important role in
financial markets.

The bohemian wave function is as of the form
\begin{eqnarray}
\psi(q,t)=R(q,t) \exp(i\frac{S(q,t)}{\hbar}),
\end{eqnarray}
where $R(q,t)$ is
the amplitude of the wave function  and $S(q,t)$ is its phase.
Substituting $\psi(q,t)$ in the Schr$\ddot{o}$dinger's equation before
differentiating twice, two equations for $R$ can be obtained:
\begin{eqnarray}
\frac{\partial{R^2}}{\partial t} &+& \frac{1}{m}\frac{\partial({R^2\frac{\partial s}{\partial q}})}{\partial q}=0,
\nonumber \\
\frac{\partial S}{\partial t}&+&\frac{1}{2m}(\frac{\partial S}{\partial q})^2+(V-\frac{\hbar^2}{2mR}\frac{\partial^2
R}{\partial q^2})=0.
\end{eqnarray}

The second term in Eq. 2 contains an additional potential term besides the
classical potential ($V$) which was named by Bohm as a quantum potential
\cite{Bohm,Choustova1}.
\begin{eqnarray}
 U=\frac{\hbar^2}{2mR}\frac{\partial^2 R}{\partial q^2}.
\end{eqnarray}
The hard financial conditions of
markets such as natural resources and hard relations between traders
can be described by the classical potential \cite{Choustova1}.

Choustova used the quantum potential as a method for describing the
mental factors or psychological aspects of the markets
\cite{Choustova,Choustova1,Choustova2,Choustova3}. Traders at the
financial market behave stochastically due to free wills of
individuals, where the combinations of a huge number of free wills
lead to an additional stochastic term at the market that could not
be described by classical potential.

In the following section, by applying Bohmian approach to financial
markets and finding the quantum potential of the market returns, we
try to find the psychological aspects of the markets. This means
that the today market return is entangled by the previous returns,
where by considering this relationship, we obtain fine information
about the market conditions.

\begin{figure}[t]
\includegraphics[width=14cm,height=11cm,angle=0]{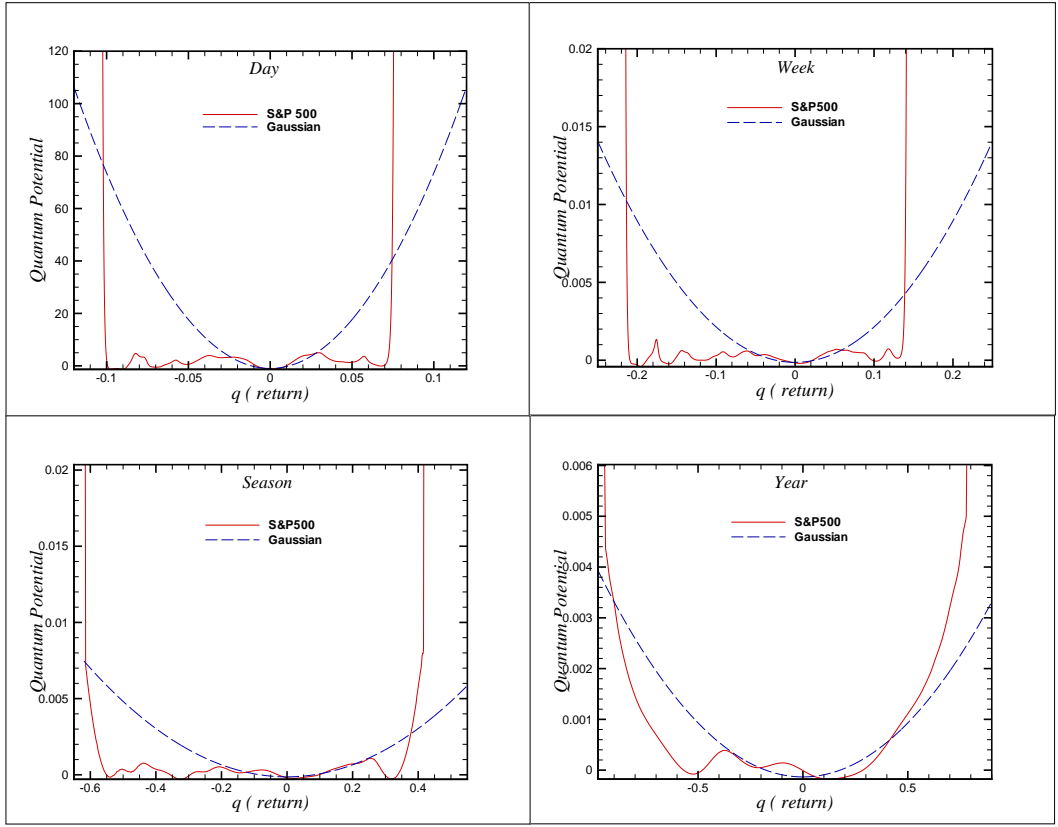}
\caption{the quantum potential of return time series of S\&P500 index in
different time scales and their comparison with the white
noise with the same variance in each panel.} \label{fig1}
\end{figure}

\begin{figure}[t]
\includegraphics[width=11cm,height=9cm,angle=0]{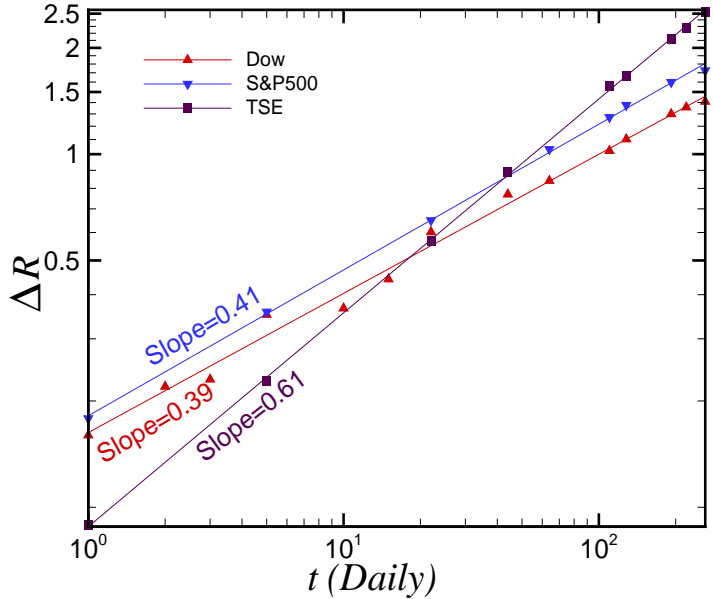}
\caption{The comparison of the scaling behaviors of the quantum
potential for emerge and mature markets.} \label{fig2}
\end{figure}

\begin{figure}[t]
\includegraphics[width=11cm,height=9cm,angle=0]{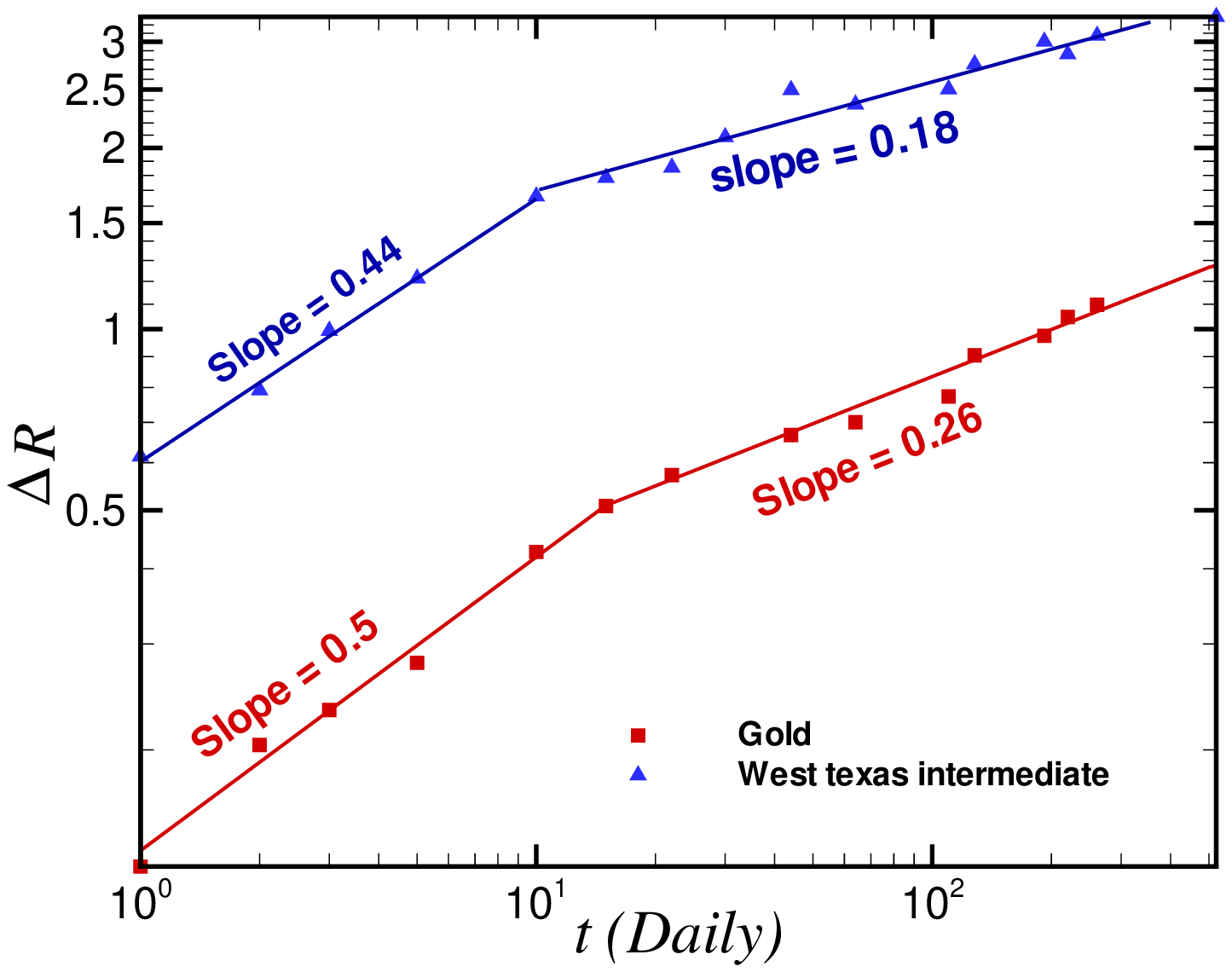}
\caption{The scaling behaviors of the quantum potential for
commodity markets.} \label{fig3}
\end{figure}


\section{Practical use of Bohmian quantum}

The quantum potentials of different markets have been found by
assuming $m=1$ based on the statistical inference of the probability
density functions of the markets. Fig.1 shows the quantum potential
of return time series of S\&P500 index in different time scales. As we
can see in daily scale, the maximum probable increase in returns is
almost 7 percent. On the other hand, the maximum probable decrease
in returns is about $- \%10$. In order to compare these data
with real data, it is interesting to mention that there are three
days that their returns are much more greater than the maximum limit
($\%7$) and only there is one day (Black Monday- 19 Oct 1987)
that its return is smaller than $- \%10$. As we can see for
each market, there are two potential limits where the probability of
occurrence of any returns outside these limits is very low. In
essence, these limits act as boundary to prevent the change of
return time series. Based on the previous formulas, the quantum
potential has an inverse relation with the probability distribution
functions of the time series of returns. By comparing the behavior
of a white noise with SPX index, it is obvious that between the two
limits, the real market has much higher probability for occurrence.
This phenomenon is stemmed from the fat-tail of the distributions of
real markets. On the other hand, outside of these limits, the
probability of occurrences for the white noise is much more than the
SPX. In Fig . 2, we have depicted the differences between two
extreme limits for some mature and emerging markets during different
time scales. It is obvious that the slope of the scaling behaviors
in the mature market indices (SPX and DJIA) vs. scale approximately
is about 0.4. On the other hand, the value of the slope for the
emerge market index (TSE) is greater than 0.5. This behavior reminds
us about the behavior of the Hurst exponent in emerging and matures
markets. As shown in previous studies, the Hurst values for emerge
and mature markets are greater and less than 0.5 respectively.

 Fig .3 shows that there is a cut-off in the slopes of the commodity
markets (oil and gold). The slopes of these markets have two
different values where they are the same as mature markets. Before
the cut-off points, in short scales the important factor that
affects the structure of the markets is the competing nature of the
firms in terms of supply and demand. So, the commodity markets in
this region have the same nature as the mature markets. But, on the
other hand in the large scales the most influential parameters that
shape the behavior of the markets are the political and macro
subjects. So, the behavior of the commodity markets in large scales
is different from the short scales.

\section{Conclusion}

In this paper, we have used the Bohmian quantum approach to analyze the financial markets. By using this approach we can describe the behavioral aspects of the market indices such as the coupling structure of markets returns during different times in terms of a quantum potential. One of the important findings of this work is that the probable returns for each market lay in specific boundaries during different time scales.

Also, by comparing the real market with a white noise, it is obvious that a real fat-tail distribution has a much more probability for occurrence within the boundaries. On the other hand, for a white noise, the occurrence for any return could be made possible by increasing the quantum potential, while for the real market this issue is out of the question.

By estimating the differences between maximum and minimum returns for some markets, it is found that the quantum potentials of the return series possess scaling behaviors. Also, the slopes of this scaling behavior in the mature and commodity markets are less than 0.5 and on the other hand, the slope of the emerge market (TSE) is greater than 0.5.

In addition, the cut-off observed in the curves of the commodity markets indicates the different efficiency of local and global effects.

The finding would remind us about the behavior of the Hurst exponent, that was shown in previous work for mature markets, where the values where greater than 0.5 for emerge markets and less than 0.5 for mature markets.

Based on the conclusions of this work, the Bohmian quantum approach is a useful technique for analyzing the financial markets.

\section{Acknowledgement}
The authors would like to thank Sara Zohoor and Dr. S. Vasheghani
Farahani for helping to edit the manuscript.


\end{document}